\documentclass[twocolumn]{svjour3}          
\smartqed  
\usepackage{graphicx}
\begin{document}

\title{Topological Phases emerging from Spin-Orbital Physics
}


\author{Wojciech Brzezicki \and Mario Cuoco \and Filomena Forte \and Andrzej M. Ole\'s
}


\institute{Wojciech Brzezicki \and Mario Cuoco \and Filomena Forte \at
             CNR-SPIN, IT-84084 Fisciano (SA), Italy;\\
             Dipartimento di Fisica \textquotedblleft{}E. R. Caianiello\textquotedblright{},\\
             Universit\'a di Salerno, IT-84084 Fisciano (SA), Italy
           \and
           Andrzej M. Ole\'s \at
             a.m.oles@fkf.mpg.de \\
             Max Planck Institute for Solid State Research,\\
             Heisenbergstrasse 1, D-70569 Stuttgart, Germany;\\
             Marian Smoluchowski Institute of Physics, Jagiellonian University,
             prof. S. \L{}ojasiewicza 11, PL-30348 Krak\'ow, Poland
}

\date{Received: date / Accepted: date}

\maketitle

\begin{abstract}
We study the evolution of spin-orbital correlations in an inhomogeneous
quantum system with an impurity replacing a doublon by a holon orbital
degree of freedom. Spin-orbital entanglement is large when spin
correlations are antiferromagnetic, while for a ferromagnetic host we
obtain a pure orbital description. In this regime the orbital model can
be mapped on spinless fermions and we uncover topological phases with
zero energy modes at the edge or at the domain
between magnetically inequivalent regions.

\end{abstract}

\section{Introduction}

Transition metal oxides are fascinating materials where several degrees
of freedom (i.e., spin, orbital, charge, etc.) couple and, from a
theoretical point of view, need to be treated on equal footing in order
to provide reliable predictions. In undoped $3d$  Mott insulators
large on-site Coulomb interactions localize electrons and the coupling
between transition metal ions is controlled by a low-energy spin-orbital
superexchange introduced first by Kugel and Khomskii \cite{Kug82}.
As for spins, orbital degrees of freedom have a quantum character and
can drive strong fluctuations which end in destroying long range order
\cite{Fei97} or lead to exotic novel types of magnetic order
\cite{Brz12}. However, such cases are rare in $e_g$ systems and,
typically, long range order in both spin and orbital sector develops in
perovskite lattices \cite{Ole05}, with the corresponding correlations
following the Goodenough-Kanamori rules \cite{Goode}. A well known
example is the spin-orbital order in LaMnO$_3$ \cite{Kov10}, with
different energy scales for spin and orbital order \cite{Sna16}.
However, there are numerous deviations from these rules caused either
by superexchange on non-linear bonds \cite{Gee96}, or by lattice
frustration such as for instance in LiNiO$_2$ \cite{Rei05}, or by
spin-orbital entanglement \cite{Ole06}, or, finally, by the presence of
next nearest neighbor hopping \cite{Hyper}.
In $t_{2g}$ systems, orbital superexchange has leading contributions
with SU(2) symmetry along a given cubic direction thus orbital
fluctuations are much stronger \cite{Kha05} than in $e_g$ and a
spin-orbital liquid emerging from intrinsic frustration is more likely
to occur \cite{Nor08,Cha11,Karlo}.
On the other hand, ordered states may be even stabilized by orbital
fluctuations \cite{Kha01} as for instance in LaVO$_3$ \cite{Fuj10} and
Ca$_2$RuO$_4$ \cite{Cuo06} --- in this latter case spin-orbit coupling
also plays a role \cite{Fiona}. Quantum fluctuations and spin-orbital
entanglement \cite{Ole12} are of great importance in this class of
materials and may lead to novel phenomena as superconductivity in the
pnictides driven by competing symmetries at orbital degeneracy
\cite{Nic11}, or spectacular topological structure of the excited
states in the one-dimensional (1D) SU(2)$\otimes$XY model \cite{Brz14},
or, finally, dimerised phases \cite{Imre}.

Doping of Mott insulators adds another charge degree of freedom in
spin-orbital systems and leads to several remarkable phenomena.
Recently short range charge-density wave called stripe phase was
reported in doped cuprates \cite{Cam15}. It has been suggested that
the critical charge, orbital, and spin fluctuations near the quantum
critical point provide the pairing interaction \cite{Bia00}.
As in doped cuprates, the holes doped in $t_{2g}$ orbitals may be
mobile due to three-site terms \cite{Dag08} or self-organization in
stripe phases \cite{Wro10}. However, the formation of orbital
molecules makes 1D insulating zigzag states kinetically more
favorable than metallic stripes \cite{Prl15}. Insulating state is also
found \cite{Ave15} when holes are confined near charge defects in
Y$_{1-x}$Ca$_x$VO$_3$~\cite{Hor11}.

In contrast, neutral defects in spin-orbital systems lead to orbital
dilution (with a local increase of spin to $S=\frac32$) \cite{Brz15}
and the changes in spin-orbital order \cite{Brz16}, or charge dilution
\cite{Brz17} (with invariant spin $S=1$ states). These phenomena are
distinct from the orbital dilution in cuprates where holes remove
simultaneously spin and orbital degree of freedom \cite{Tan09}.
The $t_{2g}$ systems with charge dilution are unexplored yet --- they
will likely play a major role in future functional materials and,
possibly, in novel electronic devices.
The purpose of the paper is to investigate the
consequences of \textit{charge dilution} in a $t_{2g}$ system due to
the substitution of a $d^4$ by a $d^2$ transition metal ion. Such type
of doping allows to uniquely design a spin-orbital correlated
environment with an orbital degree of freedom having an inequivalent
\textit{charge} character. Indeed, for $d^2$ and $d^4$ valence
configurations, the empty orbital (i.e., \textit{holon}) and the doubly
occupied state (i.e., \textit{doublon}) set the orbital degree of
freedom, respectively.
As an experimental motivation we mention, among the various emergent
phenomena and the many possible hybrid oxides which could be designed,
that:
($i$) dilute Cr doping for Ru reduces the temperature of the
orthorhombic distortion, induces ferromagnetic (FM) order and anomalous
negative thermal expansion in Ca$_2$Ru$_{1-x}$Cr$_x$O$_4$
(with $0<x<0.13$) \cite{Qi10}, and
($ii$) Mn-substituted single crystals of Sr$_3$Ru$_{2-x}$Mn$_x$O$_7$
rapidly drive an unusual metal-insulator transition and $E$-type
antiferromagnetic (AF) order at low doping \cite{Mes12}.
The theoretical search for the consequences of
\textit{holon}-\textit{doublon} substitution is performed for a 1D ring
and we analyze both spin and orbital correlations around the charge
defect. We give reasons why the FM regime is well designed to
search for topological aspects of the present model.

\section{Spin-orbital physics and charge dilution}

\begin{figure}[t!]
\begin{center}
\includegraphics[width=.9\columnwidth]{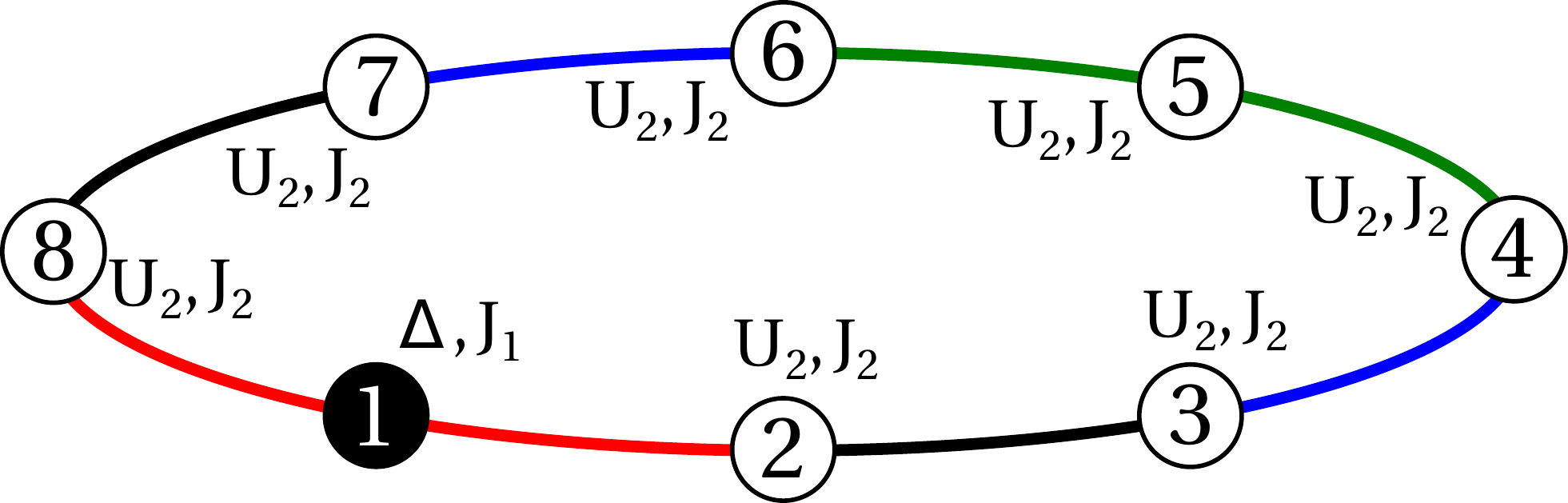}
\end{center}
\caption{
Artist's view of a ring of length $L=8$ containing seven $d^4$
host atoms $i=2,\dots,8$ with parameters $\{U_2,J_2\}$ and one charge
impurity $d^2$ at site $i=1$ (filled circle),
with parameters $\{\Delta,J_1\}$.
The color convention for the bonds $\langle i,i+1\rangle$ is the same
as in Figs. 2-4.
}
\label{fig:0}
\end{figure}

We consider a 1D ring made of $d^4$ transition metal ions in the
insulating regime, with one $d^2$ charge defect, see Fig.
\ref{fig:0}. The physics of the undoped system is governed by a
spin-orbital superexchange model which is equivalent, through an
electron-hole transformation, to that introduced for vanadates
\cite{Kha04}. It depends on two Kanamori's parameters: the
intraorbital Coulomb element $U_2$ and Hund's exchange $J_2$ for
$t_{2g}$ electrons \cite{Ole05}, responsible for the high spin states
with spin $S=1$ at the host $d^4$ ions. In the regime of strong
electron interactions we obtain a spin-orbital model with spin $S=1$ at
every site and an orbital degree of freedom described by a pseudospin
$T=\frac12$. Since we work in one dimension and with $t_{2g}$ orbitals,
we select the cubic axis $c$ with the active orbitals \cite{Kha01}:
$\left|a\right\rangle\equiv\left|yz\right\rangle$ and
$\left|b\right\rangle\equiv\left|xz\right\rangle$.

However, the situation becomes less familiar when some of the $d^4$
ions are substituted by the $d^2$ ones. In the regime of low doping,
all the bonds will be either between
two $d^4$ ions, called host bonds, or between $d^2$ and $d^4$ ions
around an impurity site --- these we call hybrid bonds. The superexchange
Hamiltonian for both kinds of bonds has a generic form (all the bonds are
along the cubic axis $c$ in the 1D chain) \cite{Brz17},
\begin{equation}
{\cal H}_{b}=J_{b}\sum_{\langle i\rangle}
\left\{K_{i,i+1}^{(b)}\vec{S}_{i}\cdot\vec{S}_{i+1}+Q_{i,i+1}^{(b)}\right\},
\label{eq:Ham}
\end{equation}
where the label $b=\{{\rm 0,h}\}$ stands for the type of bond and the
operators $\left\{K_{i,i+1}^{(b)}\right\}$ and
$\left\{Q_{i,i+1}^{(b)}\right\}$ act in the orbital space at two sites,
${\{i,i+1\}}$.
These operators differ fundamentally for the hybrid and host bonds,
i.e., for the host they take the U$(1)$ symmetric form of,
\begin{eqnarray}
\label{Khost}
K_{i,i+1}^{(0)}\!&=&A_K\tau_i^z\tau_{i+1}^z
+B_K\left(\tau_i^x\tau_{i+1}^x+\tau_i^y\tau_{i+1}^y\right)+C_K, \\
\label{Qhost}
Q_{i,i+1}^{(0)}\!&=&A_Q\tau_i^z\tau_{i+1}^z
+B_Q\left(\tau_i^x\tau_{i+1}^x+\tau_i^y\tau_{i+1}^y\right)+C_Q,
\end{eqnarray}
whereas for the hybrid bonds the symmetry is lowered,
\begin{eqnarray}
\label{Khyb}
K_{i,i+1}^{\rm (h)}\!&=&\!D_K\tau_i^z\tau_{i+1}^z
+E_K\tau_i^x\tau_{i+1}^x+F_K\tau_i^y\tau_{i+1}^y+G_K,  \\
\label{Qhyb}
Q_{i,i+1}^{\rm (h)}\!&=&\!D_Q\tau_i^z\tau_{i+1}^z
+E_Q\tau_i^x\tau_{i+1}^x+F_Q\tau_i^y\tau_{i+1}^y+G_Q,
\end{eqnarray}
Here, $\tau_i^{\alpha}$ are the Pauli operators describing doublon/holon
fluctuating between $|a\rangle$ and $|b\rangle$ orbitals. They are
defined by the Pauli matrices $\vec{\sigma}$ as
\begin{equation}
\vec{\tau}_{i}=\big(\begin{array}[t]{cc}
a_{i}^{\dagger} & b_{i}^{\dagger}\end{array}\big)\cdot\vec{\sigma}\cdot
\big(\begin{array}[t]{cc}a_i^{} & b_i^{}\end{array}\big)^T,
\end{equation}
where hardcore boson operators $a^{\dagger}_i$ and $b^{\dagger}_i$
create holon or doublon in the orbitals $|a\rangle$ and $|b\rangle$,
respectively.

Coefficient $J_{b}$ in the spin-orbital model (\ref{eq:Ham}) is a
superexchange constant and is given by $J_0=4t_0^2/U_2$ and
$J_{\rm h}=2t_{\rm h}^2/\Delta$ (note that the excitations which
provide the main contribution on the hybrid bond go in one direction
only), where $t_b$ ($b=0,{\rm h}$) is a hopping amplitude along
host-host or impurity-host bond, and $U_2$ is the Hubbard interaction
for the host ($d^4$) atoms. $\Delta$ is a typical excitation energy
scale in the virtual process $d^2_id^4_j\Rightarrow d^3_id^3_j$
(or charge transfer energy) given by,
\begin{equation}
\Delta\equiv I_{e}+2U_{1}-3U_{2}-6(J_{1}+J_{2}).
\label{Delta}
\end{equation}
Here $U_1$ and $J_1$ are Hubbard and Hund's interactions at the $d^2$
impurity site. $I_{e}$ is the energy mismatch of the electronic levels
at two ions; since $\Delta$ must be positive and relatively
large, this implies that $I_{e}>U_i$ must be the largest energy scale
in the system. The coefficients $A_{K(Q)},\dots,C_{K(Q)}$ and
$D_{K(Q)},\dots,G_{K(Q)}$ in Eqs. (\ref{Khost})--(\ref{Qhyb}) are
numerical constants depending on microscopic parameters of the ions:
$A_{K(Q)},\dots,C_{K(Q)}$ depend only on host's parameter $\eta_0$ and
$D_{K(Q)},\dots,G_{K(Q)}$ both on host's and impurity's Hund's exchange
$\eta_1$ and $\eta_2$, where
\begin{equation}
\label{para}
\eta\equiv\frac{J_2}{U_2},      \hskip .7cm
\eta_1\equiv\frac{J_1}{\Delta}, \hskip .7cm
\eta_2\equiv\frac{J_2}{\Delta}.
\end{equation}
All these $\eta$'s measure the relative strength of Hund's exchange
with respect to typical excitation energy --- in case of host bonds
it is $U_2$ whereas for hybrid bonds it is $\Delta$ (\ref{Delta}).
The exact functional forms of these coefficients are complicated and
will be reported elsewhere.

The properties of the host and hybrid bonds are the following:
a single host's bond is always FM in spin and AF in orbital sector
because of an orbital singlet which is formed on a bond \cite{Kha01}.
This however is not stable when there are more than one bond --- for
a longer 1D system as for $L=8$ chain considered here we find AF spin
correlations for low $\eta$ (\ref{para}) turning FM in a high $\eta$
limit. The case of a hybrid bond is much simpler:
despite the complicated form of the Hamiltonian (\ref{eq:Ham}) it
always gives AF spin correlations accompanied by FM
$\langle\tau^z_i\tau^z_{i+1}\rangle$ orbital correlations.
Because of these intrinsic difference between host and hybrid bonds it
is essential to check the ground state properties of a finite system
with single impurity, see Fig. \ref{fig:0}.

\begin{figure}[t!]
\begin{center}
\includegraphics[width=8cm]{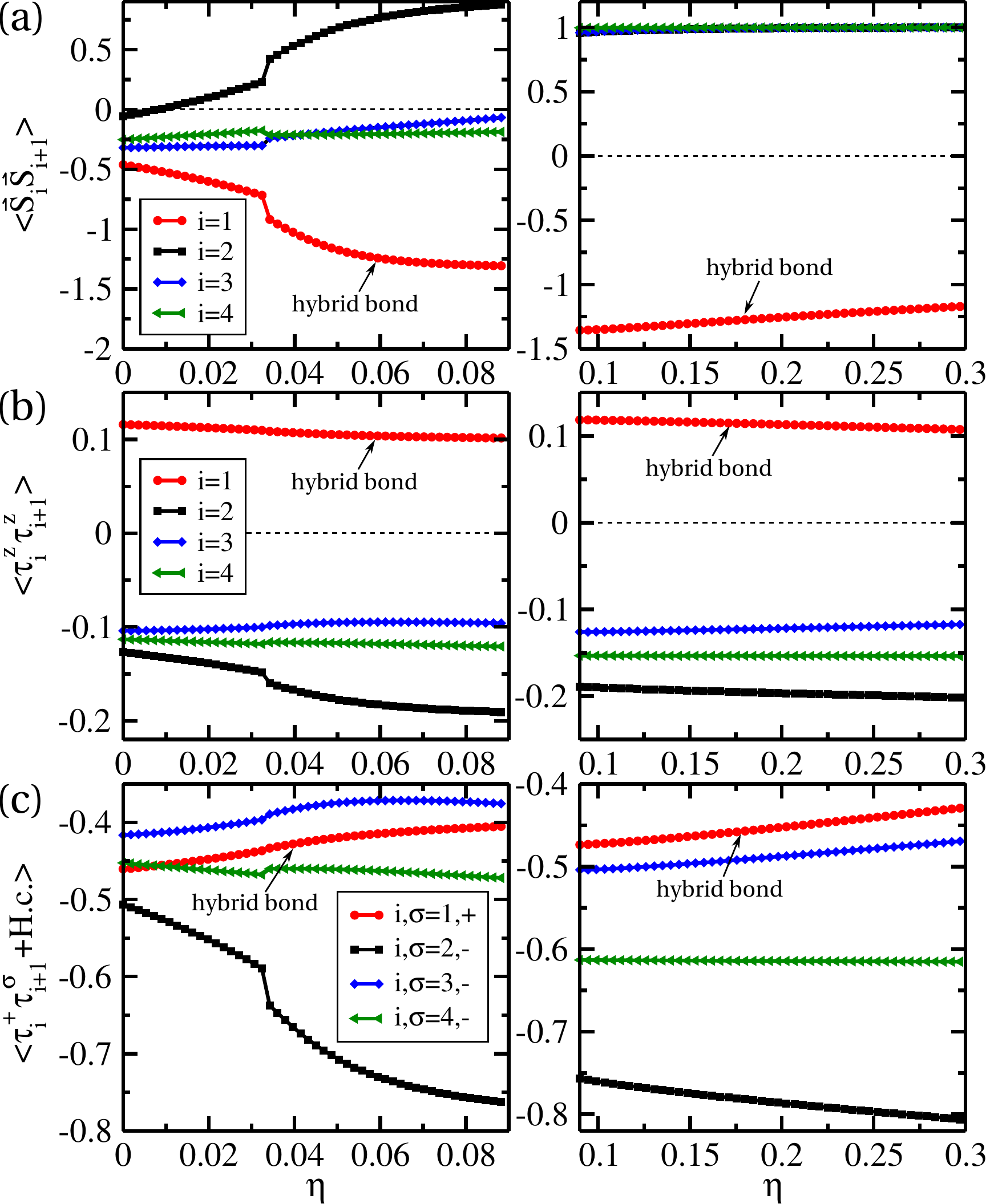}
\end{center}
\caption{
Ground-state spin and orbital correlations on the bonds for a closed
chain of length $L=8$ shown in Fig. 1. The computation is performed by
means of exact diagonalization as a function of $\eta$. Left column
--- AF host with small $\eta$, right column --- FM host with larger
$\eta$. Rows:
(a) spin $\langle\vec{S}_i\vec{S}_{i+1}\rangle$ correlations,
(b) orbital $\langle\tau^z_i\tau^z_{i+1}\rangle$ correlations, and
(c) orbital
$\left\langle\tau_i^+\tau_{i+1}^-\right\rangle$
($\left\langle\tau_i^+\tau_{i+1}^+\right\rangle$) correlations
for the host/hybrid bond.
Parameters: $J_{{\rm 0}}=1$, $J_{{\rm h}}=2$, $\eta_1=\eta_2=1$.
}
\label{fig:1}
\end{figure}

In Fig. \ref{fig:1} we show the ground state spin and orbital
correlations obtained for a closed chain of $L=8$ sites with a single
$d^2$ impurity, see Fig. \ref{fig:0}. The results are shown
as functions of $\eta$ for fixed values of $J_0$, $J_{{\rm h}}$,
$\eta_1$, and $\eta_2$ which weakly influence
the overall behavior. Due to translational invariance one finds
four inequivalent bonds, see Fig. \ref{fig:0}.
There are two regimes:
($i$)~AF with total spin ${\cal S}=2$
($\left\langle\sum_i S^z_i\right\rangle=2$) for $\eta<0.09$,
($ii$)~FM with ${\cal S}=6$ ($\left\langle\sum_i S^z_i\right\rangle=6$)
for $\eta>0.09$ (however the hybrid impurity bonds are always AF).
In the AF regime at $\eta=0$ all the spin correlations are~AF, but
a level crossing occurs at $\eta=0.033$ where the magnetic moment
delocalizes from the impurity to its two neighbors, remaining nearly
constant within these three sites.

Surprisingly, for increasing $\eta<0.09$ the spin correlations between
second and third neighbors of impurity become soon FM, due to
spin-orbital entanglement, but the remaining spin correlations are AF.
In the FM regime all the host bonds have almost saturated FM spin
correlations, $\simeq+1$, while they tend to the classical value of
$-1$ for increasing $\eta$ on hybrid bonds, see Fig. \ref{fig:1}(a).
The orbital $\langle\tau^z_i\tau^z_{i+1}\rangle$ correlations behave
more regularly; they are AF for host bonds and FM for hybrid bonds in
both regimes of $\eta$, see Fig. \ref{fig:1}(b). For the off-diagonal
orbital correlations we define the conventional $\tau^{\pm}_i$
operators as $\tau^{\pm}_i\equiv\frac12(\tau^x_i\pm\tau^y_i)$
(here $\tau^{x(y)}_i$ are normalized to $\pm 1$). It turns out that
$\langle\tau^+_i\tau^-_{i+1}\rangle$ correlations are significant only
for the host bonds and $\langle\tau^+_i\tau^+_{i+1}\rangle$
only for hybrid bonds and they are always AF, see Fig. \ref{fig:1}(c).

\begin{figure}[t!]
\begin{center}
\includegraphics[width=1.\columnwidth]{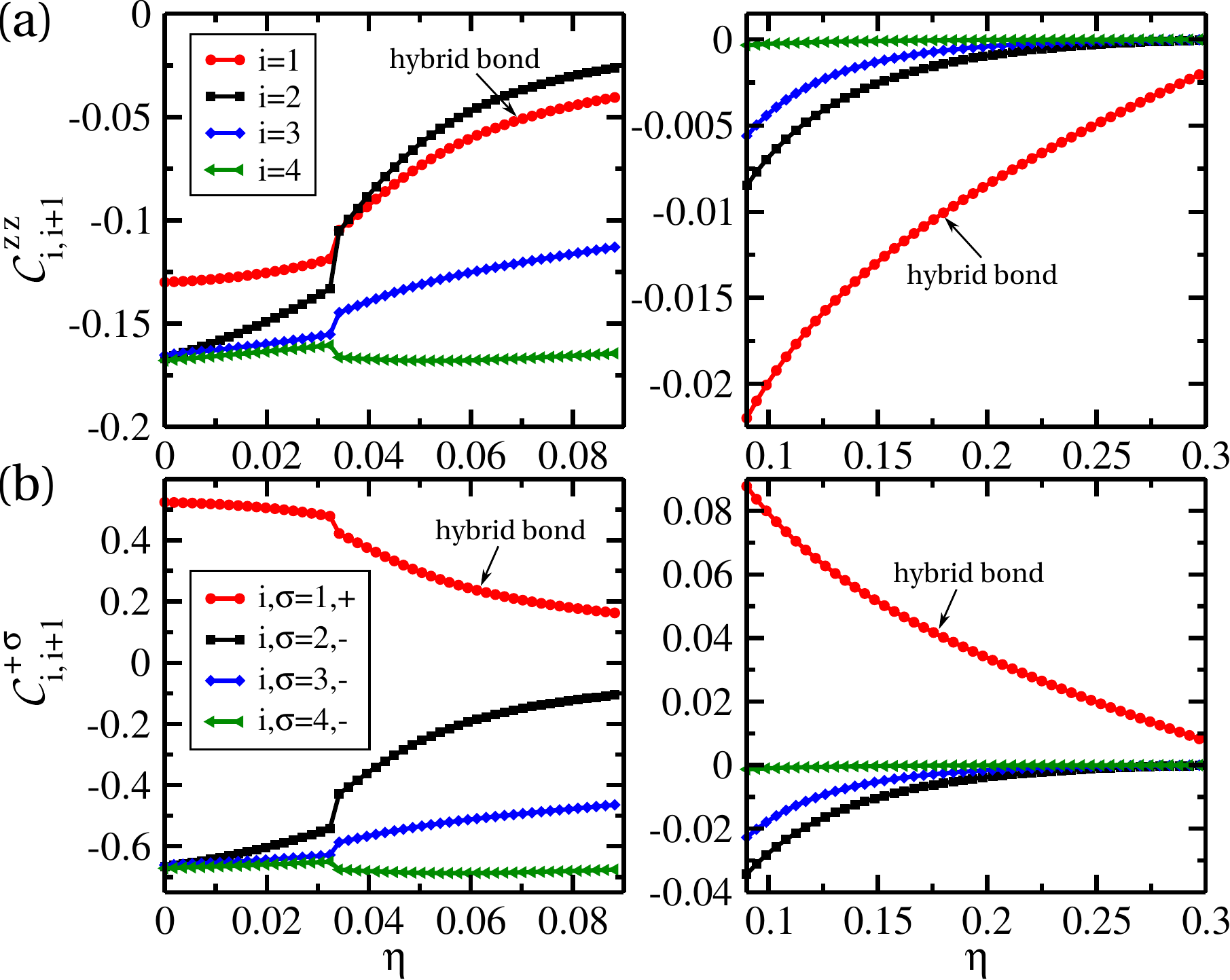}
\end{center}
\caption{
Ground state spin-orbital covariances ${\cal C}_{i,i+1}^{\alpha\beta}$
on the bonds for a closed chain system of length $L=8$ (Fig. 1),
obtained via exact diagonalization as functions
of $\eta$ for orbital correlations:
(a)~$\left\langle\tau^z_i\tau^z_{i+1}\right\rangle$;
(b)~$\left\langle\tau_i^+\tau_{i+1}^-\right\rangle$
   ($\left\langle\tau_i^+\tau_{i+1}^+\right\rangle$) for the host (hybrid)
   bonds.
Left (right) column --- AF (FM) host with small (large) $\eta$.
Parameters: $J_{{\rm 0}}=1$, $J_{{\rm h}}=2$, $\eta_1=\eta_2=1$.
}
\label{fig:2}
\end{figure}

To investigate the spin-orbital entanglement we introduce covariances
for the various correlators,
\begin{eqnarray}
\label{Czz}
{\cal C}_{i,i+1}^{zz}\!&=&
\left\langle{\vec S}_i{\vec S}_{i+1}\tau^{z}_i\tau^{z}_{i+1}\right\rangle \!-\!
\left\langle{\vec S}_i{\vec S}_{i+1}\right\rangle
\left\langle\tau^{z}_i\tau^{z}_{i+1}\right\rangle,
\\
\label{C++}
{\cal C}_{i,i+1}^{+\sigma}\!&=&
\left\langle{\vec S}_i{\vec S}_{i+1}\tau^{+}_i\tau^{\sigma}_{i+1}\right\rangle \!-\!
\left\langle{\vec S}_i{\vec S}_{i+1}\right\rangle
\left\langle\tau^{+}_i\tau^{\sigma}_{i+1}\right\rangle
+{\rm H.c.},
\end{eqnarray}
with $\sigma=\pm$. In Fig. \ref{fig:2} we show the spin-orbital
covariances in the AF and FM regime. One finds that both longitudinal
(${\cal C}_{i,i+1}^{zz}$) and transverse (${\cal C}_{i,i+1}^{+\sigma}$)
covariances are large in the AF regime. Moreover, as one could expect,
they are much lower at higher $\eta>0.1$ when the host spin
correlations are FM, while they tend to zero as $\eta$ increases, see
Fig. \ref{fig:2}. Interestingly, the transverse covariance for the
hybrid bond is positive (${\cal C}_{i,i+1}^{++}>0$) in the entire regime
of parameters which suggests that double orbital excitations are strong
on hybrid bonds. Thus, we conclude that the factorization into spin and
orbital operators is a good approximation only in the FM regime and for
this case we set the spin-spin correlations as equal to $\pm 1$ for the
host-impurity bonds.

\section{ Topological states in the orbital model }

Factorization of spin and orbital degrees of freedom is allowed in the
FM regime and leads to an effective orbital-only Hamiltonian,
\begin{eqnarray}
\label{orbi}
H_{i,j}^{{\rm 0}}&=&\frac{1}{4}\,J_{{\rm 0}}\,
\frac{1}{1-3\eta}\,\vec{\tau}_i\vec{\tau}_{j},    \nonumber\\
H_{i,j}^{{\rm h}}&=&J_{{\rm h}}\left(
A_{xx}\tau^x_i\tau^x_j+A_{yy}\tau^y_i\tau^y_j+A_{zz}\tau^z_i\tau^z_j\right),
\end{eqnarray}
for the host and hybrid bonds, respectively. This purely orbital
Hamiltonian can be mapped on spinless fermions by the
Jordan-Wigner transformations.

\begin{figure}[t!]
\begin{center}
\includegraphics[width=1.\columnwidth]{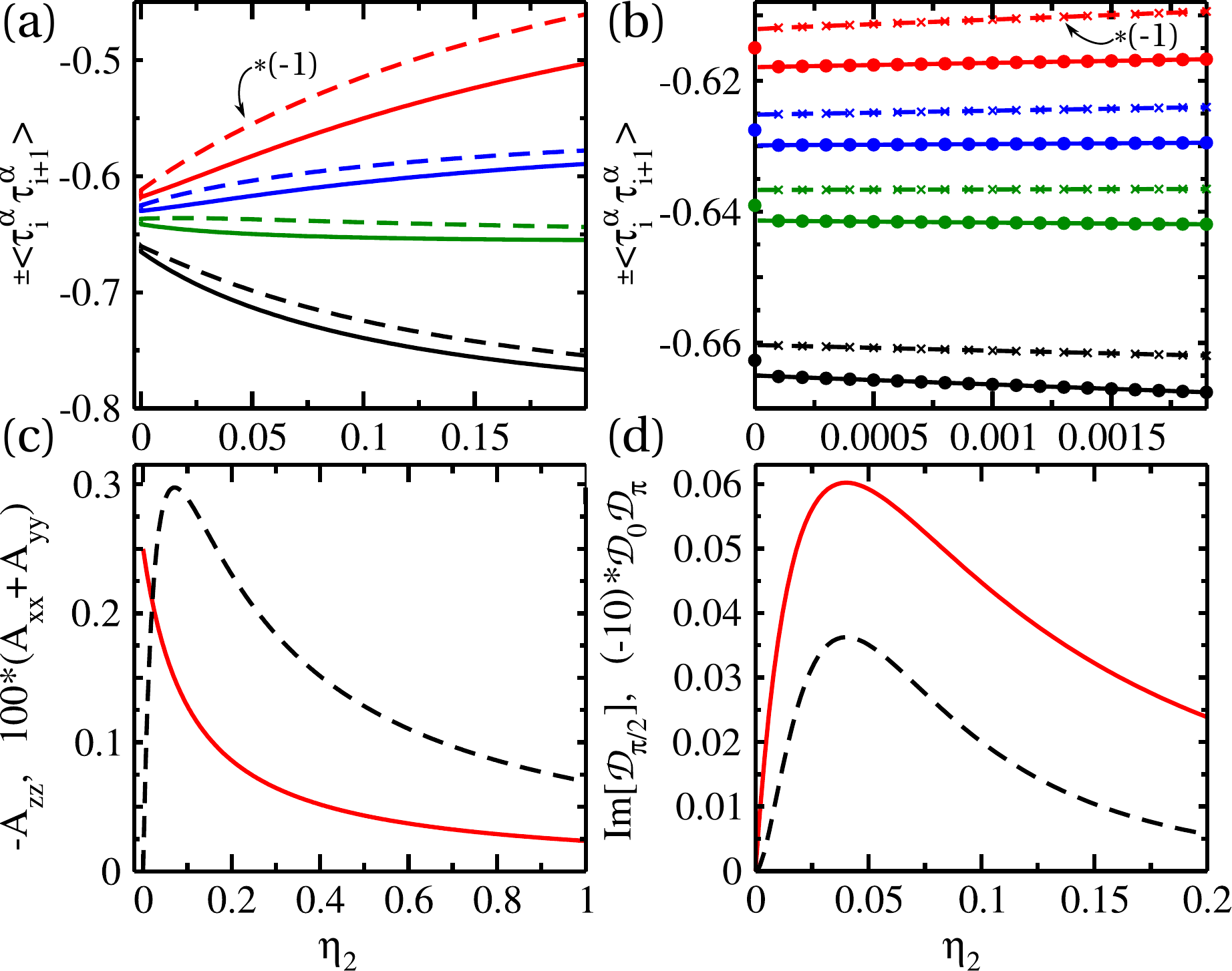}
\end{center}
\caption{
Orbital model results obtained in Hartree-Fock (Eq. \ref{orbi})
in the regime of FM host as a function of $\eta_2=J_2/\Delta$:
(a)~bond correlations $\langle\tau_i^x\tau_{i+1}^x\rangle$ and
$\pm\langle\tau_i^y\tau_{i+1}^y\rangle$ (solid and dashed lines),
with color convention as in Fig. 1;
(b)~magnified view of (a) for $\eta_2\to 0$;
(c) orbital couplings $A_{zz}$ and sum $A_{xx}+A_{yy}$
(solid and dashed) for hybrid bonds, and
(d)~topologically relevant quantities, ${\rm Im}{\cal D}_{\pi/2}$
and ${\cal D}_0{\cal D}_\pi$ (solid and dashed).
Parameters: $J_{{\rm 0}}=1$, $J_{{\rm h}}=2$, $\eta_1=4$, $\eta=0.2$.}
\label{fig:3}
\end{figure}

For symmetry reasons we find an exact relation $A_{zz}\equiv -A_{xx}$,
and we also get that $A_{xx}$ and $A_{yy}$ almost compensate each
other so their sum $A_{xx}+A_{yy}=\delta$ has a relatively small
amplitude. It is however important to point out that $\delta\not=0$
because in the representation of Jordan-Wigner fermions $\delta$ is
proportional to the pairing amplitude and its finite value can induce
a topological non-trivial state. All the $\{A_{\alpha,\alpha}\}$
coefficients are functions of $\eta_1$ and $\eta_2$, while we find that
the dependence on $\eta_1$ is very weak. Thus, we fix $\eta_1=4$
($\eta$ is already fixed as $\eta=0.2$) and we show the behavior of the
$A_{\alpha,\alpha}$  coupling in Fig. \ref{fig:3}(c). We note that
$\eta_2=0$ is a high symmetry point where $A_{\alpha,\alpha}\equiv0.5$
for any $\eta_1$.

Hence, by means of the Hartree-Fock decoupling we deal with
fermion-interaction term $\langle\tau^z_i\tau^z_{i+1}\rangle$ in a
self-consistent manner and we obtain the bond
$\left\langle\tau^{\alpha}_i\tau^{\alpha}_{i+1}\right\rangle$ orbital
correlations for a periodic (and infinite) system with one $d^2$
impurity per every $L=8$ sites. We find that in the present parameter
regime the $\langle\tau^z_i\tau^z_{i+1}\rangle$ vanish and one gets
only the kinetic terms $\langle\tau^x_i\tau^x_{i+1}\rangle$ and
$\langle\tau^y_i\tau^y_{i+1}\rangle$. For the host bonds they are all
AF while for the hybrid ones the $xx$ correlations are AF and $yy$ ones
are FM, see Fig. \ref{fig:3}(a).
Interestingly, we obtain a discontinuous transition at $\eta_2=0^+$
between anisotropic and isotropic phases --- the difference between
$xx$ and $yy$ correlations is triggered by any finite $\eta_2$, see
Fig. \ref{fig:3}(b). Finally, at finite $\eta_2$ one always gets
a regime with a non-trivial topological phase with respect to the
Jordan-Wigner fermionic representation.

Indeed, the fermionic Hamiltonian in the momentum space is given by a
matrix ${\cal H}_k$ that belongs to the BDI Altland-Zirnbauer class
\cite{Altland}. Thus, it can have a non-trivial $Z$ topological number.
The topological invariant can be determined by looking at ${\cal H}_k$
in the eigen-basis of the chiral symmetry where it consists of two
anti-diagonal blocks $u_k$ and $u_k^{\dagger}$. The determinant of
$u_k$, ${\cal D}_k\equiv\det{u_k}$, is a complex number which yields
a non-trivial topological number if it winds around the $(0,0)$ point
in the complex plane as $k$ changes from $0$ to $2\pi$.
In present case this happens if:
($i$) the imaginary part of $u_{\pi/2}$ is non-vanishing and
($ii$) the determinants ${\cal D}_0$ and ${\cal D}_{\pi}$ have opposite
signs.
In Fig. \ref{fig:3}(d) we observe that indeed these conditions hold as
long as $\eta_2>0$.

\begin{figure}[t!]
\begin{center}
\includegraphics[width=1.\columnwidth]{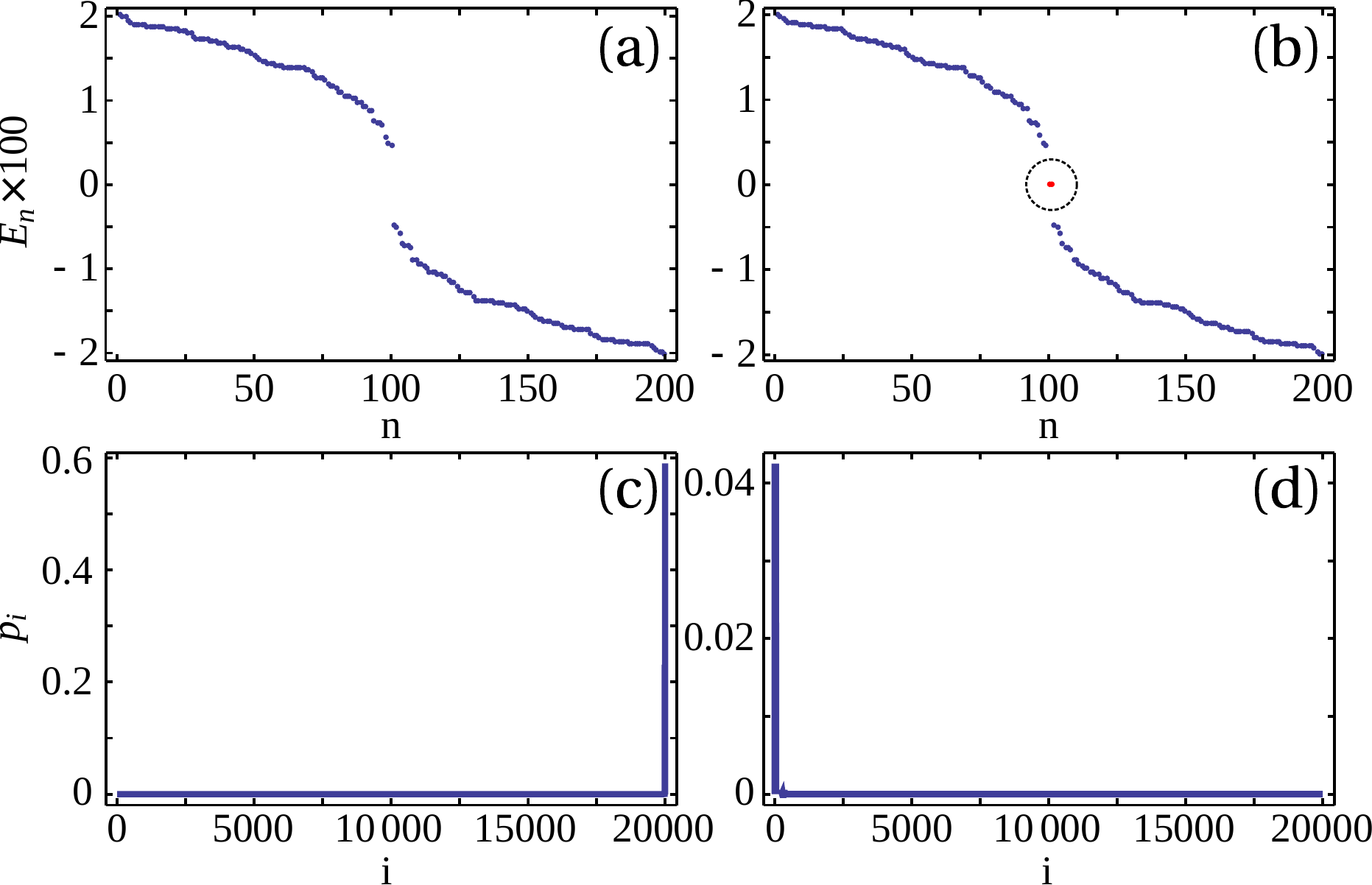}
\end{center}
\caption{
Energy spectra and edge states for a fully disordered 1D chain (\ref{chain})
of size $L\!=\!20000$ and $N\!=\!200$ random impurities for:
(a)~a~closed chain, and
(b) an open chain, exhibiting two zero energy states in the gap.
The occupation probabilities $\{p_i\}$ for the Majorana zero-energy
states in the gap at:
(c)~the right chain edge, and
(d)~the left chain edge.
}
\label{fig:5}
\end{figure}

Recently, the topological phase diagram of a 1D tight-binding model
of spinless electrons with an inhomogeneous distribution of pairing
centers has been investigated \cite{Top17}. The Hamiltonian includes
inhomogeneities generated by diluted pairing centers with a given
distribution profile in the unit cell of length $L$.
For a periodic configuration with momentum $k$ we get,
\begin{eqnarray}
\label{chain}
{\cal H}&=&\!\sum_{p=1\atop k}^{L}\!
\left\{t_{p}^{} c_{kp}^{\dagger}c_{k,p+1}^{}
\!+\!\Delta_p^{} c_{kp}^{\dagger}c_{-k,p+1}^{\dagger}\!+\!\textrm{H.c.}
\!+\!\mu_{p}^{} c_{kp}^{\dagger}c_{kp}^{}\right\}\!,\nonumber \\
\end{eqnarray}
with $c_{L+1,k}\equiv e^{ik}c_{1,k}$ and
$\{t_{p},\Delta_{p}\}$ being the nearest neighbor hopping and on-bond
pairing amplitudes. There, we have found the topological invariant that
can be generally expressed in terms of the physical parameters for any
pairing center configuration \cite{Top17}.

Here, we emphasize the occurrence of edge states and present the
spectra around zero energy for a closed and open system, see Figs.
\ref{fig:5}(a) and \ref{fig:5}(b). We note that for an open system
there are two zero-energy states appearing in the gap. These are
Majorana end modes that arise as a consequence of the bulk-boundary
correspondence in a topologically non-trivial configuration.
In Figs. \ref{fig:5}(c) and \ref{fig:5}(d) the spatial occupation
probabilities $p_i$ for the two zero energy states are explicitly shown
in order to confirm their degree of localization on the right/left
edges of the 1D chain.
We also point out that the modification of the kinetic term with the
inclusion of long-range hopping is expected to lead to multiple
Majorana end modes both in spinless \cite{DeGot} and spinfull $p$-wave
superconducting chains \cite{Merca}.

\section{Discussion and Summary}

In conclusion, we have studied a one-dimensional hybrid $d^2$-$d^4$
system with a single $d^2$ impurity in a $d^4$ spin-orbital correlated
host. Remarkably, the exact diagonalization analysis allows to single
out regimes for which the orbitals and spins can be factorized if the
host configuration is FM. By this decoupling one finds interacting
orbital pseudospins exhibiting fully isotropic exchange for the host
bonds and fully anisotropic for the hybrid ones. A Jordan-Wigner
transformation and Hartree-Fock decoupling allow, then, to map the
system on non-interacting fermions and to find topological non-trivial
states. Unexpectedly, a topological non-trivial state occurs for any
finite value of $J_2$, i.e., the amplitude of the Hund's coupling at
the host's ions. For a long chain we explicitly demonstrate that
Majorana-like modes occur at the edge of the system. We argue that
inhomogeneous topological patterns \cite{Marra} can be achieved in
the present spin-orbital scenario with Majorana modes occurring,
for instance, at the boundary of the FM region if the impurities
drive a magnetic configuration that has alternating FM with AF domains.
\vskip .3cm

\noindent\textbf{Acknowledgments}
Open access funding provided by Max Planck Society.
W.B. acknowledges support by the European Union's Horizon 2020
research and innovation programme under the
Marie Sk\l{}odowska-Curie grant agreement No. 655515.
We acknowledge support by Narodowe Centrum Nauki
(NCN, National Science Center, Poland),
Project No.~2016/23/B/ST3/00839.

\end{document}